\documentclass[runningheads]{svmult} 
\usepackage{makeidx}   
\usepackage{graphicx}  
\usepackage{subeqnar}  
\usepackage{multicol}  
\usepackage{cropmark} 
\usepackage{physprbb}  
\begin{document} 
\title*{Self-Interacting Dark Matter} 
\toctitle{Self-Interacting Dark Matter} 
%
%
\titlerunning{Self-Interacting Dark Matter} 
%
\author{Benjamin D.\ Wandelt\inst{1} \and Romeel Dav\'e \inst{2} \and
Glennys R. Farrar\inst{3} \and Patrick C. McGuire\inst{4,5}
David N.\ Spergel\inst{2} \and Paul J.\ Steinhardt\inst{1}  }
\authorrunning{Wandelt et al.} 
%
%
\institute{Department of Physics, Princeton University, Princeton NJ
08544, USA \and Department of Astrophysics, Princeton University,
Princeton NJ 08544, USA \and 
Department of Physics, New York University, New York, NY 10003  USA
\and Zentrum f\"ur Interdisziplin\"are Forschung, 
Universit\"at Bielefeld, Bielefeld, Germany 33615 \and  Steward Observatory
and Department of Physics, University of Arizona, Tucson, AZ 85721
}  
\maketitle              

\begin{abstract} 
Spergel and Steinhardt have recently proposed the concept of dark 
matter with strong self-interactions as a means
to address numerous discrepancies between
observations of dark matter halos on subgalactic scales and the 
predictions of the standard collisionless dark matter picture. 
We review the motivations for this scenario and discuss 
some recent, successful numerical tests.
We also discuss the possibility that the dark matter interacts
strongly with ordinary baryonic matter, as well as with itself.
We present a new  analysis of the experimental constraints and
re-evaluate the  allowed range of cross-section and mass.  

\end{abstract}  

\section{Why Interacting Dark Matter Particles?} 
%

Cosmological models with a mixture of roughly 35\% collisionless
cold dark matter  such as axions, WIMPs, or any other candidate
interacting   through the weak and gravitational forces only, and 
65\% vacuum energy or quintessence match observations
of the cosmic microwave background and large scale structure on 
extra-galactic scales with remarkable accuracy
\cite{BahcallETAL,WangETAL}.  However, a number of discrepancies have
arisen between numerical simulations and observations on subgalactic
scales.\cite{Moore1} We will address these now in turn. 

{\bf Halo density profiles.} 
Simulations of collisionless cold dark matter halos (CCDM)
\cite{Moore1,NFW,kravtsov,klypin} consistently show 
cuspy halo density profiles similar to the 
generalized Navarro--Frenk--White \cite{NFW} (hereafter NFW) profile 
\begin{equation}
\rho(r)=\rho_0\frac{r^{3}_s}{r^\alpha(r+r_s)^{3-\alpha}}
\label{NFWprofile}
\end{equation}
where $r_s$ is a scale radius, which is related to the virial
radius $r_h$ of the halo via the concentration parameter
$c=r_h/r_s$.  This profile has the feature that it diverges as
$\rho\sim r^{-\alpha}$ as $r\rightarrow 0$. 

The exact value for $\alpha$ found in simulations is a 
matter of some debate -- NFW
originally found $\alpha\simeq 1$, while more recent, higher
resolution simulations find somewhat steeper divergences, $\alpha\simeq
1.5$. This is unlikely to be an artifact; if anything, resolution and
discreteness effects lead  to simulations underestimating the
steepness of the central density profile.  Biases can occur when
fitting (\ref{NFWprofile}) to simulated halos. For example, the fact
that there is a finite number of particles per halo make the choice of
the halo center somewhat arbitrary. Any departures from the ``true''
center will give the illusion that the  density profile is less cuspy
than it actually is.

Observationally, gravitational lensing in clusters 
and the presence of disks in galactic halos make it possible to infer
their mass distributions. Observations of low surface 
brightness galaxies (LSB) and dwarfs, chosen for their small
mass-to-light ratios, indicate cores of galactic
halos with shallow density profiles \cite{flores,deBlok,dalcanton}. 
Similarly, there is some
evidence that the central regions of cluster halos not associated with
luminous galaxies 
appear have flat density profiles ($\rho =r^0$ as $r\rightarrow 0$) \cite{tyson}.

More recently, preliminary results from  high quality measurements of rotation
curves in edge-on LSBs with thin disks
\cite{dalcantontalk} indicate a best fit of $\alpha\simeq 0.5$, more
mildly divergent than and inconsistent with CCDM  profiles.

Note that only exquisitely
well-measured rotation 
curves simultaneously constrain {\em both} the concentration parameter
$c$ and the slope $\alpha$ \cite{unsmeared}. In practice, observations
do not resolve these parameters independently.
This means that the problem of  halo 
shapes can be rephrased in terms of $c$; observed halos tend to have
smaller concentration ($c\simeq 6-8$) than simulated ones ($c\simeq
20$). The crucial point is that whether one focuses on slope or
concentration,  simulations and observations do not agree.


 
{\bf Excess of small scale structure.} 
One of the most striking features
of halos in high resolution CCDM simulations is that they are heavily
populated with small subhalos or subclumps. The scarcity of observed
satellite galaxies \cite{MooreSub} could be interpreted as another failure of the CCDM
paradigm on small scales. 

Even if there are mechanisms which inhibit star
formation in small clumps and thereby leave them dark,
a large population of clumps may heat the galactic disk and even
endanger its stability \cite{MooreSub,TothOstriker,Weinberg}.

This is a delicate issue; there clearly is halo substructure in  galaxies
(satellites, dwarfs, high velocity clouds). Any proposed solution must
reduce the number of satellites but guard against wholesale smoothing of
the density field on small scales.

{\bf Tully-Fisher relation.} Further evidence for less centrally
concentrated halos comes from the failure of CCDM models to account
for the zero-point of the Tully--Fisher relation. 
Mo and Mao \cite{MoMao} explain the surprisingly tight relation
between the luminosity of disk galaxies and
their maximum  rotation speed in terms of the gravitational
interactions between galaxy disk and halo. Their model requires the
presence of a dark halo with concentration $c\simeq3$, much less  than $c\simeq20$ which
is found in simulations. Significantly, this pertains to all disk
galaxies -- not just LSBs and dwarfs.

{\bf Bar stability.}
In high surface brightness spiral galaxies, the persistence of bars
implies low-density cores \cite{deBattistaSellwood}.

It is conceivable that these discrepancies are due to 
problems with current simulations, the quality of present observations
or the omission of important
astrophysical processes in the models. However, if any one of them
persists, they may be an indication that the dark matter is not
collisionless. 

In this paper we argue that {\em all} of the 
discrepancies listed above may be resolved if the dark matter
has strong self-interactions.  If this proves to be 
the case, the study of fine-scale structure on subgalactic (kiloparsec)
scales, a subject that has not been given much attention by particle
physicists in the  past, could prove to be a remarkably powerful and
precise probe
of the properties of dark matter.




\section{The Physics of Self-Interacting Dark Matter Halos} 

Consider a CCDM halo.
The particles in its center have
small rms speeds (are ``cold''). As subclumps sink into the
larger halo, their outer, hotter,  less tightly bound parts are
tidally stripped and 
remain in the outer parts of the halo. If the cold centers of the
clumps survive, they join the cold cusp of the halo. Because of the lack of
two-body interactions (the gravitational potential is smooth) there is
no heat transport into the cold CCDM cusp. 

Virialized CCDM halos are not in thermal equilibrium.
When interactions are present, however, the second law of thermodynamics
requires that heat be transferred into the cold
center from the outer halo to approach equilibrium.  Initially, the
increased rms speeds spread out the halo cusp into a smoother
core. At the same time as the inner part is heated, the outer part is
cooled, the cooler material moves in, is scattered and mixed,
so that eventually  the halo becomes isothermal at late times. 
Particles will be ejected due to two-body interactions, the halo
loses heat and gravothermal collapse sets in, moderated by ongoing
accretion of small subclumps, which continue to heat the halo.

An important fact to keep in mind is that the  heat conduction
coefficient does not have a simple dependence on the cross
section. A traditional analysis {\em \`a la} Chapman and Enskog
\cite{Liboff} expands about thermal
equilibrium which is  brought about by frequent interactions.
In this  regime, where the mean free path of the particles is much
shorter than any characteristic length scale of the system, the heat conduction
{\em decreases} with increasing cross-section,
$\kappa(r)\propto \frac{\pi k_B}{\sigma}
\sqrt{\frac{\langle v^2(r)\rangle}{ \pi}}$.

As we will see below, the interactions discussed here are 
constructed to be in a different regime altogether. Here, the mean
free path is of the order of the scale of the halo.  In this regime,
namely the  small cross-section limit, heat
transport {\em increases} as the cross-section increases. For this
reason a numerical implementation of self-interacting dark matter is
non-trivial and cannot be achieved by solving the standard equations
of cosmological hydrodynamics. Also, because of the non-monotonic
dependence on cross-section, a simple interpolation
between the case of  collisionless dark matter and the hydrodynamic
limit is not sufficient.  

The presence of scatterings has another important effect: the
particle velocities become randomized. The random component of
particle velocities in
a collisionless fluid can be highly correlated, leading to non-spherical 
velocity ellipsoids \cite{BinneyTremaine}. 
Collisions erase these correlations, randomizing the motions, which
leads to more spherical velocity ellipsoids. The resulting lack of a preferred
direction should isotropize the inner parts of halos where scattering
is important.

Spergel and Steinhardt \cite{SpergelSteinhardt}  postulated a
self-interaction between dark matter particles with strength which
encompassed the range 
\begin{equation}
s=\frac{\sigma_{DD}}{M}=8\times 10^{-25}- 1\times10^{-23}\, 
{\rm cm}^2\,{\rm GeV}^{-1}=0.5-6\, {\rm cm}^2\,{\rm g}^{-1}
\label{sigma}
\end{equation}
many orders of magnitude stronger than for CCDM
particles. (The range quoted here is narrower than originally proposed
in \cite{SpergelSteinhardt} because of additional constraints
discussed below). A convenient dimensionless quantity measuring the
strength of the 
interaction at the solar circle is the  optical depth $\tau=s/s_{\rm
solar}$ where $s_{\rm solar}= 
28.4\,{\rm cm}^2/{\rm g}$. Unit optical depth  corresponds to a mean free
path of order the diameter of the solar orbit $\lambda_{8\,{\rm
kpc}}=16\,{\rm kpc}$ at the densities of the solar neighborhood
($\rho_{8\,{\rm kpc}}\simeq10^5 \rho_{\rm crit}\simeq 7\times
10^{-25}\,$g/cm$^3$). One can then estimate the size of the
optically thick region  $R_{opt}\simeq 8\tau\,{\rm kpc} $.  

The range  of $s$
originally quoted in \cite{SpergelSteinhardt} was chosen to cover a
span of optical depths straddling unity.  
On the optically thin side, subhalos are destroyed by 
{\it spallation} in which  particles  are
scattered out of the subhalo one-by-one 
almost every time there is an impact with a
hotter particle from the surrounding halo. 
On the optically 
thick side, subhalos are destroyed by {\it evaporation} in
which the subhalo  absorbs all the kinetic energy 
from nearly every  incoming 
hot particle of  the surrounding halo, heats up and expands. 
Either mechanism suffices for the purpose of destroying dwarf halos
in galaxies. However,
a problem with evaporation is that it is so efficient that 
galaxy halos in massive clusters probably  cannot survive, which conflicts
with observations.  Eq.~(\ref{sigma}) takes account of this 
problem by shaving off the optically thick regime from the 
range of $s$ found in \cite{SpergelSteinhardt}.

\section{Numerical Studies}

To summarize, the qualitative  effects of introducing 
self-interacting dark matter (SIDM) in the regime of cross-sections in
(\ref{sigma}) are: 
\begin{itemize}
\item {\em Heat transport}  leading to modified halo profiles and core
collapse on time scales larger than the Hubble time.
\item {\em Isotropization of
the velocity tensor} in dense regions and hence more  spherical shapes of
the inner parts of halos.
\item {\em Limited destruction of halo substructure} through spallation.
\end{itemize}
Intriguingly, the observations we listed in the introduction suggest
precisely this sort of  behavior.

\begin{figure}[t] 
\begin{center}
\includegraphics[width=\textwidth]{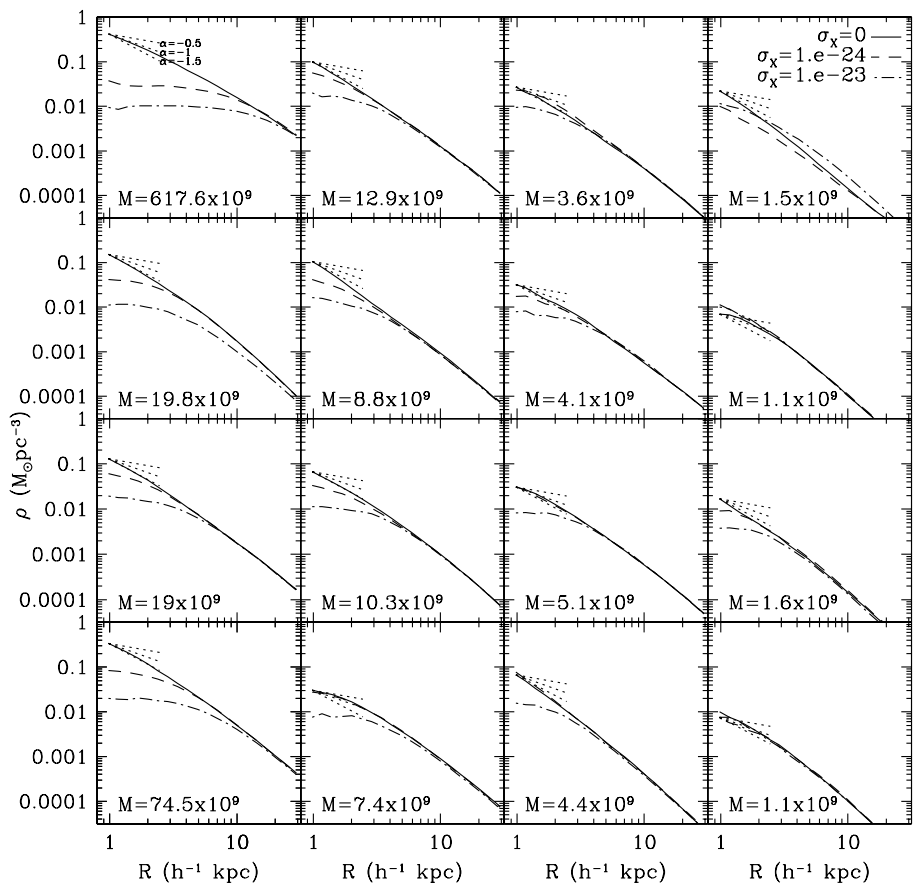} 
\end{center}
\caption[]{Comparison of halo profiles for non-interacting (solid) and
self-interacting dark 
matter with cross-sections $\sigma_{DD} =10^{-24}\,$cm$^2$GeV$^{-1}$ (dashed) and $\sigma_{DD}=10^{-23}\,$cm$^2$GeV$^{-1}$ (dot-dashed). 
Halos are chosen from a descending range of masses, with the four
largest halos in our simulation shown in the leftmost column
Three short rays corresponding to logarithmic slopes $0.5$, $1$ and
$1.5$ are shown to guide the eye in the inner part of the halo
} 
\label{fig:halocomp} 
\end{figure}   

This has motivated several authors
\cite{Hannestad,Burkert,MooreColl,Yoshida,KochanekWhite,Yoshida2} 
to study the effects of self-interactions numerically. 
Apart from  simulations which are done in the hydrodynamic limit
\cite{MooreColl,Yoshida}, all of these
find that SIDM halos contain smooth cores.
However, the 
longevity of these cores and the time-scale for gravitational
collapse are still a matter of debate. 

In particular, Kochanek and White simulate isolated halos and find
fast evolution 
towards gravothermal collapse for higher mass halos
\cite{KochanekWhite}.  {\em Caveats} of their simulations include
that   1) they begin with steep, high-density profiles which are
already cuspy before the interactions are turned on,
which leads to 
fast evolution because the central density starts from artificially
high values,
2) they do not include accretion, which
ought to  moderate 
gravothermal collapse and 3) the cross-section
they simulate are high. For a Milky Way type galaxy ($R_0\simeq10
R_{\rm disk}
\simeq35\,kpc$, $M_0\simeq7\times10^{11}M_\odot$) their simulations
cover a range from 
$s\simeq2.5-83\,{\rm g}/{\rm cm}^2$. They observe core collapse for
cross-sections above our range  (\ref{sigma}).
Burkert \cite{Burkert}
has performed   simulations comparable to Kochanek and White's, 
and his results suggest significantly longer lived cores.
Yoshida {\em et al.}
\cite{Yoshida2} compute the halo profile of a cluster from
cosmological initial conditions. They, too, find
large cores which persist until today.

For our simulations, we modified the freely 
available GADGET tree N-body code \cite{GADGET} to include scatterings
between dark matter particles using Monte Carlo techniques similar to 
\cite{KochanekWhite}. The simulations begin with
cosmological initial conditions in the Lambda model currently favored
by the majority of observations ($\Omega_0=1=\Omega_m+\Omega_\Lambda$,
$\Omega_\Lambda=0.7$). We follow the
evolution of dark matter, 
represented by $128^3$ particles in a box with 4$h^{-1}$\,Mpc on the
side and  1$h^{-1}$\,kpc spatial resolution. Periodic boundary
conditions  ensure that we model the 
continuing accretion of halos correctly. 

We simulate interaction strengths of  $s\simeq
0.06$, $0.6$, $6$ and $60$\,cm$^2$/g. Out of these the smallest value
of $s$ produced results that are
barely distinguishable from collisionless dark
matter. The 
largest cross-section considered significantly suppresses the formation of
halos, and those that do form are much too large and diffuse.
But for the two intermediate cases our galactic
halos consistently developed constant density cores.

%


\begin{figure}[t] 
\begin{center}
\includegraphics[width=.618\textwidth]{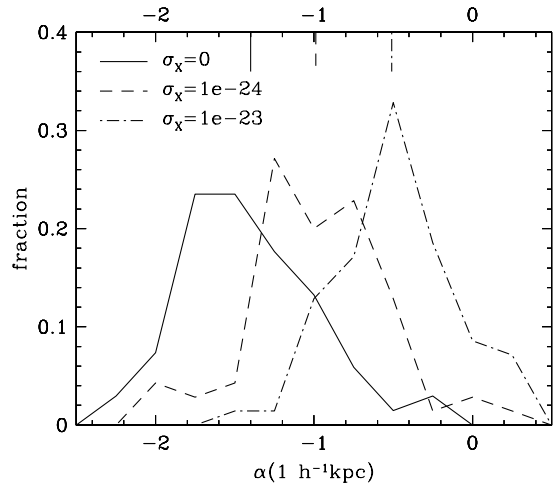} 
\end{center}
\caption[]{Histogrammed distribution of inner halo slopes $\alpha$ for
non-interacting (solid) and self-interacting dark matter with
cross-section $\sigma_{DD} =10^{-24}\,$cm$^2$GeV$^{-1}$ (dashed) and
$\sigma_{DD}=10^{-23}\,$cm$^2$GeV$^{-1}$ (dot-dashed). 
We only include halos with more than 1000 particles, corresponding to masses
greater than $3.6\times 10^9$\,$M_\odot$} 
\label{fig:alphadist} 
\end{figure}

Figure \ref{fig:halocomp} compares the profiles of  halos from
three of our simulations which assume different cross-sections
but which are otherwise identical.
It is clear that for intermediate cross-sections, within  the range
suggested by Spergel and 
Steinhardt, core collapse has not set in. Consistent with our previous
discussion of the optically thin regime, we find evidence that
the stronger the self-interactions are, the smoother the halos become.

Figure \ref{fig:alphadist} shows the histograms of inner slopes
measured for all halos with more than 1000 particles in each of the 3
simulations. It is clear that in the regime we are exploring,
increasing the cross-section smoothes the halo profiles.
 
These and other results from our simulations  will be discussed in 
more detail in a forthcoming 
publication \cite{Dave}.


\section{Could Dark Matter Consist of an Exotic, Neutral, Stable Hadron?} 
\label{sec:DMbaryon}

If self-interactions are responsible for resolving the issues of the halo profile and galactic substructure,
then the ratio of the dark matter-dark matter cross-section $\sigma_{DD}$ to the mass  of the dark matter particle must
lie in the range given in Eq.~(\ref{sigma}).  Curiously, the ratio for ordinary hadrons, such as 
neutrons and protons, lies in roughly this same range, ${\cal O}(10^{-23} \, cm^2/GeV)$ at low energies.
One natural possibility is that the dark matter belongs to the hidden sector of a supergravity or superstring 
model with hidden gauge interactions analogous to ordinary gluon exchange.     
A second possibility
is that the dark matter is composed of
exotic hadrons ---  neutral, stable  particles 
 which interact through the  strong force with ordinary 
quarks and gluons as well as with themselves.
In this case, 
the ratio of scattering cross-section to mass for self-scattering ($s$)
might be comparable to the ratio for dark matter-nucleon scattering.
We should emphasize that the self-interaction
proposal does not require that there be strong interactions between dark matter and ordinary hadrons; it suffices
if the dark matter interacts strongly with itself.
However,  the fact that the cross-section required to solve the halo profile problem is so similar to the  cross-section for the familiar
strong interaction makes it impossible to resist
considering  the possibility.  

If the exotic hadron is neutral and  stable and  does not
radiate photons or other massless mesons, then 
it will act like dissipationless, collisional dark matter.  Possible examples include the 
quark-gluino bound state, the $S0$ (in some models, the lightest supersymmetric 
stable
particle with odd $R$-parity and unit baryon number) \cite{Farrar,steinhardt}
or strangelets \cite{Wilczek}.  Alternatives are non-hadronic
exotics which might also be arranged to have 
the desired value of $\sigma_{DD}/m$, such  as
gauge singlet mesons \cite{Bertolami}, and
$Q$-balls \cite{Kusenko}.  Of these, the
first three naturally have values of $s= \sigma_{DD}/m$ which lie in the desired range, Eq.~(\ref{sigma}).  Another condition
is that the exotic hadron must  not bind to light nuclei. Otherwise, the particle would efficiently
convert hydrogen and helium to exotic nuclei  during nucleosynthesis and would be found plentifully.
  Binding to light nuclei may be sufficiently suppressed if the particle has quantum numbers that forbid simple
pion exchange with ordinary nuclei, as might
occur if the particle is an isospin singlet.  
Yet a further constraint is that the cosmic abundance must 
account for the observed matter density.
  If the particle has similar mass and interactions to ordinary 
  hadrons, perhaps a natural explanation 
can be found for why the cosmic energy density in exotic hadrons 
is comparable that of ordinary 
hadrons.   

One may suppose that the notion of  exotic hadron is easily discounted. 
Surely the particle is already
 ruled out by existing cosmological and astrophysical constraints or 
 by existing
 accelerator experiments or  
dark matter detectors or searches for exotic nuclei. 
Surprisingly, the story is not nearly so simple.  For example, if the
particle is strongly interacting and similar to a neutron 
in most characteristics, it evades all currently
published constraints known to the authors. During nucleosynthesis, its kinetic energy density is too small to
affect the cosmic expansion rate and, if it does not bind 
to light nuclei, it has no effect on ordinary nuclear
abundances or evades terrestrial searches for unusual nuclei.  Only an
undetectably small fraction of the particles are captured by stars, such as the Sun, so they have no significant 
effect on stellar evolution, supernovae, etc.  The scattering 
cross-section for dark matter on 
ordinary protons and neutrons $\sigma_{Dp}$ would presumably be close
 to the dark matter-dark matter 
cross-section,
$\sigma_{DD}$, so one concern is that halo dark 
matter collides with ordinary baryons in the disk,
creating excess high energy cosmic rays and destabilizing the disk.  
We find that these considerations 
do not rule out the exotic hadron scenario, either. 
In cases where the  mass and interaction strength 
is similar to the 
more abundantly produced neutrons, it is difficult to exclude 
them  with accelerator experiments.

Hence, one is led to consider dark matter searches. However, most dark matter searches are aimed at 
weakly interacting particles (WIMPS), such as axions and supersymmetric neutralinos, which can
penetrate the atmosphere and deep  underground unimpeded.  Consequently, the detectors are placed
at ground  level or in mines.  Strongly interacting particles interact in the upper atmosphere and will  
thermalize with the air before they reach ground level.  For a 1 GeV particle, say, the incoming 
kinetic energy is 1 keV (corresponding to a virial velocity of 300 km/s) and a local density of 0.4 GeV/cm$^3$.
If the particle is strongly interacting, then the 
average kinetic energy is an undetectable 30 meV by the time 
the particles reach ground level.

>From these examples, it is clear that the best way to search for strongly interacting particles is in balloons or
in space with detectors with little or no shielding and low energy threshold.   A decade ago, Starkman {\it et
al.} \cite{Starkman} surveyed all current observations and  experiments and  found that 
the combination of cross-section and mass suggested by Eq.~(\ref{sigma}) is
 not constrained for masses near 1 GeV or so, and 
 no new constraints have been published since -- until now.  

Figure~3 represents our new analysis of the constraints on strongly interacting dark matter \cite{steinhardt}. 
This plot assumes a spin-independent cross-section for scattering of
dark matter from nuclei with atomic mass $A$
that is $\sigma_{DA} = A^2 [m_{red}(A)/m_{red}(1)]^2
\sigma_{Dp}$, as considered by Starkman et al. \cite{Starkman}.
It is straightforward to modify the plot for spin-dependent cross-sections
or cases with different $A$-dependence, as might be expected for some
of our candidates.
The  grey strip represents the range of $\sigma_{DD}/m$ given in 
Eq.~(\ref{sigma}), which one might 
imagine is roughly similar to the dark matter-proton 
cross-section $\sigma_{Dp}/m$. 
The plot has 
three new features compared to earlier, similar exclusion plots:

\noindent
(1)~Some important corrections \cite{diss} have been applied to
constraints from older experiments (Pioneer 11 and Skylab)
compared to Starkman {\it et al.} \cite{Starkman}.   The corrections
{\it reduce} the exclusion region at large cross-section;
earlier analyses had failed to take proper account of shielding.  
We also argue that the neutron star
bound in \cite{Starkman}  does not apply in the limit of large
cross-sections \cite{steinhardt}.
The re-evaluation opens up 
a range of the grey-striped region for masses between $10^4$ to $10^{17}$~GeV.  

\noindent
(2)~We publish for the first time results of a complete analysis of the IMP7/8 cosmic ray silicon
detector satellite \cite{Mewaldt,imp78,diss} and 
the IMAX balloon-borne SIMP (strongly interaction massive
particle) detector \cite{imax1,diss,imax2}.  

\begin{figure}[t] 
\begin{center}
\includegraphics[width=\textwidth]{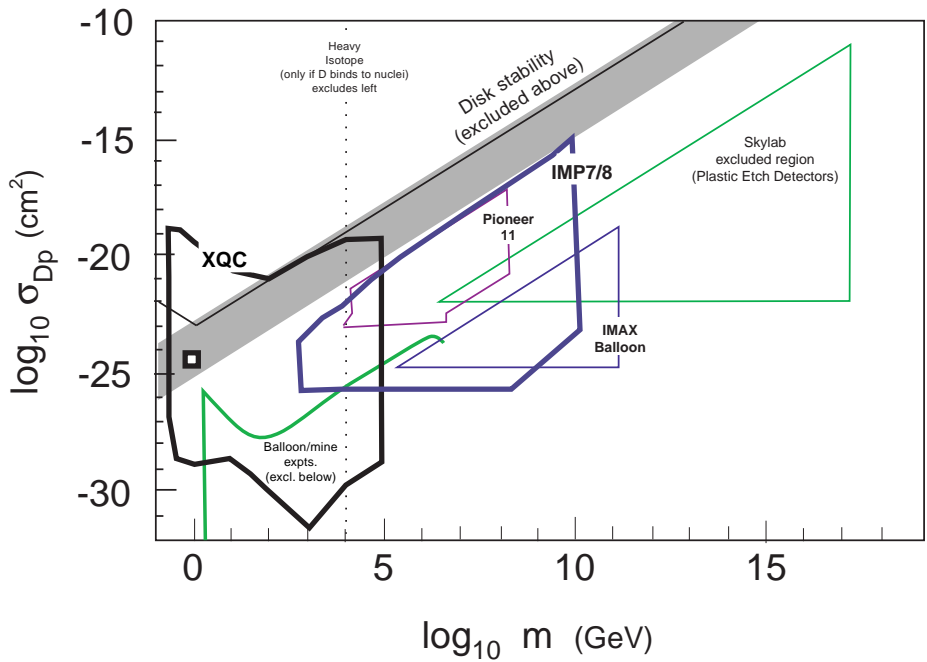} 
\end{center}
\caption[]{
Plot of the dark matter-proton scattering cross section $\sigma_{Dp}$ 
versus dark matter particle mass $m$ showing current 
experimental limits. The constraints from the XQC, IMP7/8 and 
IMAX experiments are new; the constraints from Pioneer and Skylab
experiments have been re-evaluated. The grey region shows the
range of 
dark matter-dark matter cross-section $\sigma_{DD}$  in Eq.~(2).
The strip is displayed to study the possibility that 
dark matter consists of  exotic hadrons 
in which 
$\sigma_{DD}/m$ and  $\sigma_{Dp}/m$ are comparable.
The square marks the values corresponding to a nucleon, which 
is similar to what is expected for some of the most attractive
candidates for dark matter.
The new results appear to rule out the low mass range, but
allow masses greater than $10^5$~GeV.}
\label{fig:exclusion} 
\end{figure}

\noindent
(3)~We present a provisional exclusion region for the X-ray Quantum
Calorimeter (XQC) experiment \cite{xqc},
an experiment whose first results are not  yet published \cite{xqcdata} and that was not designed with dark matter
in mind.  Remarkably, this ``inadvertent" dark matter detector provides perhaps the most
important constraint, ruling out a large, previously unconstrained  region of parameter space spanning particles
in the 0.3 GeV to $10^5$ GeV range that 
includes the $S0$ and low-mass strangelets.  (A more sophisticated
analysis\cite{forth} is likely 
to expand the exclusion region somewhat.)  The key properties of XQC
are that it 
has impressively sensitive pixels with a low energy threshold (25 eV for a substantial portion of the flight, compared to  160~keV for IMP7)
a significant detecting area, minimal atmospheric  overburden, and negligible material in front of its 1 sr
field-of-view.  The detector consists  of 33 pixels of  1 mm$^2$ area, each with an x-ray absorber
(7000 Angstroms of HgTe) deposited on 15 microns of silicon attached to
 silicon detector structure with an implanted thermometer.  Both the
HgTe and the silicon layers 
are active in detection; silicon dominates the low-mass region and
HgTe is more important for excluding 
the high mass region.
Although the detector flew on a sounding rocket at an
altitude of 170 km for only a few minutes, the low 
event rate   (probably due to x-ray sources) 
places strong constraints on dark matter candidates.  Stronger limits
can be obtained with  
subsequent flights.

\section{Conclusion}

We find that recent simulations\cite{Yoshida2,Dave}
based on 
the self-interaction proposal of Spergel and 
Steinhardt 
with  self-interactions of strength $0.1\,{\rm
cm^2/g}< s< 6\,{\rm cm^2/g}$ lend strong support
for the concept, producing results that fit
observations significantly better than standard collisionless 
cold dark matter models.
The favored dark matter candidates,
axions and neutralinos, are effectively collisionless and, hence,
are in some considerable jeopardy.  
The Spergel-Steinhardt
proposal has stimulated the interesting possibility that dark 
matter consists of particles that interact through the strong force
with ordinary matter.  Our re-evaluation of constraints 
leads us to conclude that 
the exotic hadron possibility is now ruled out for a substantial range of masses
near 1 GeV and cross-sections near $10^{-24}$\,cm$^2$, eliminating  some of the most  attractive possibilities. 
At the same time, the re-evaluation
has re-opened a region encompassing larger masses and cross-sections
previously thought to be ruled out.


\section*{Acknowledgements}
We thank Lyman Page for bringing the XQC experiment to our attention and
J. Dalcanton, N. Yoshida and D. McCammon
for sharing their
results with us prior to 
publication. 
We are also grateful to
the participants of the Princeton Monday
cosmology meetings for many useful criticisms.
RD and DNS are supported by the NASA ATP grant
NAG5-7066. DNS and BDW are
supported by the NASA MAP/MIDEX program. This work was supported in part by 
Department of Energy grant DE-FG02-91ER40671 (PJS).

%


\begin{thebibliography}{8.} 
\addcontentsline{toc}{section}{References}  
\bibitem{SpergelSteinhardt} D.N.~Spergel, P.J.~Steinhardt:
 Phys.~Rev.~Lett.~{\bf 84}, 3760 (2000)
\bibitem{BahcallETAL} N.A.~Bahcall, J.P.~Ostriker , S.~Perlmutter, and
P.J.~Steinhardt:
Science {\bf 284}, 1481 (1999)
\bibitem{WangETAL}  L.~Wang, R.R.~Caldwell, J.P.~Ostriker,  and
P.J.~Steinhardt: 
 Ap.~J.~{\bf 530},  17 (2000)
\bibitem{Moore1} S.~Ghigna, B.~Moore, F.~Governato, G.~Lake,
T.~Quinn,  J.~Stadel: {\em Density Profiles and Substructure of Dark
Matter Halos: "Converging" Results at Ultra-High Numerical Resolution }, astro-ph/9910166
\bibitem{NFW} J.F.~Navarro, C.S.~Frenk, S.D.M.~White:
 Ap.~J.~{\bf 462}, 562 (1996)
\bibitem{kravtsov} A.A.~Klypin, A.V.~Kravtsov, O.~Valenzuela, F.~Prada:   
 Ap.~J.~ {\bf 522},  82 (1999)
\bibitem{klypin} A.~Klypin:
Lecture at the Summer School ``Relativistic Cosmology: Theory and
Observations'', Como, Italy (2000)
\bibitem{flores} R.A.~Flores, J.A.~Primack:
 Ap.~J.~ {\bf 427}, L1 (1994)
\bibitem{deBlok} W.J.G.~De Blok, S.S.~Mc Gaugh:
 MNRAS  {\bf 290}, 533 (1997)
\bibitem{dalcanton} J.~Dalcanton, R.~Bernstein:
 MNRAS  {\bf 290}, 533 (1997)
\bibitem{tyson} J.A.~Tyson, G.P.~Kochanski, I.P.~Dell'Antonio:   
 Ap.~J.~{\bf 498},  L107 (1998)
\bibitem{dalcantontalk} J.~Dalcanton: Washington University, private
communication (2000)
\bibitem{unsmeared}  F.C.~van den Bosch,  B.E.~Robertson,
J.J.~Dalcanton, W.J.G.~de Blok: {\em Constraints on the Structure of Dark Matter Halos from the
Rotation Curves of Low Surface Brightness Galaxies}, astro-ph/9911372
\bibitem{miraldaescude}  J.~Miralda--Escude: {\em A Test of the Collisional Dark Matter Hypothesis from
Cluster Lensing}, astro-ph/0002050
\bibitem{MooreSub} B.~Moore,  S.~Ghigna, 
 F.~Governato, G.~Lake, T.~Quinn,  J.~Stadel, P.~Tozzi: Ap.~J.~{\bf 524}, L19 (1999)
\bibitem{TothOstriker} G.~Toth, J.P.~Ostriker:
 Ap.~J.~{\bf 389}, 5 (1992)
\bibitem{Weinberg}M.~Weinberg:
 MNRAS {\bf 299}, 499 (1998)
\bibitem{MoMao}  H.J.~Mo, S.~Mao: {\em Tully-Fisher Relation and its Implications for Halo Density
Profile and Self-interacting Dark Matter}, astro-ph/0002451
\bibitem{deBattistaSellwood} V.P.~Debattista and J.~Sellwood:
Ap.~J. {\bf 493}, L5 (1998)
\bibitem{Liboff} R.L.~Liboff:
{\em Kinetic Theory: Classical, Quantum, and Relativistic
Descriptions} (Prentice Hall, London 1990)
\bibitem{BinneyTremaine} J.~Binney, S.~Tremaine: {\em Galactic
Dynamics} (Princeton University Press, Princeton 1987)
\bibitem{Hannestad} S.~Hannestad: {\em Galactic Halos of
Self-Interacting Dark Matter}, astro-ph/9912558
\bibitem{Burkert} A.~Burkert: {\em The Structure and Evolution of Weakly Self-Interacting Cold
Dark Matter Halos}, astro-ph/0002409
\bibitem{MooreColl} B.~Moore, S.~Gelato, A.~Jenkins, F.R.~Pearce, 
V.~Quills:  Ap.~J.~Lett. {\bf 535}, 21 (2000)
\bibitem{Yoshida}  N.~Yoshida, V.~Springel, S.D.M.~White,  G.~Tormen:
{\em Collisional dark matter and the structure of dark halos}, astro-ph/0002362
\bibitem{KochanekWhite}   C.S.~Kochanek,and  M.~White: {\em A
Quantitative Study of Interacting Dark Matter in Halos}, astro-ph/0003483
\bibitem{Yoshida2}  N.~Yoshida, V.~Springel, S.D.M.~White,  G.~Tormen:
{\em Weakly Self-Interacting Dark Matter and the Structure of
Dark Halos}, astro-ph/0006134
\bibitem{GADGET} V.~Springel, N.~Yoshida, S.D.M.~White:\\
\rule{.5cm}{0cm}{\tt http://ibm-2.MPA-Garching.MPG.DE/gadget/}
\bibitem{Dave} R.~Dav\'e, D.N.~Spergel, P.J.~Steinhardt, B.D.~Wandelt:
{\em Halo Properties in Cosmological Simulations of
Self-Interacting Cold Dark Matter},
astro-ph/0006218

\bibitem{Farrar} G.R.~Farrar:    Phys.~Rev.~Lett.~{\bf 53}, 1029
(1984); Nucl.Phys.Proc.Suppl.~{\bf 62}, 485 (1998)
\bibitem{steinhardt} P.J.~Steinhardt, D.N.~Spergel, G.~Farrar, P.C.~McGuire,
B.D.~Wandelt: in preparation
\bibitem{Wilczek}  R.L.~Jaffe, W.~Busza, J.~Sandweiss, F.~Wilczek: hep-ph/9910333
\bibitem{Bertolami}  M.C.~Bento, O.~Bertolami, R.~Rosenfeld, L.~Teodoro:
{\em Self-Interacting Dark Matter and Invisibly Decaying Higgs}, 
astro-ph/0003350
\bibitem{Kusenko} A.~Kusenko and M.~Shaposhnikov: Phys.~Lett.~B {\bf
418}, 46 (1998)
\bibitem{Starkman}  G.D.~Starkman, A.~Gould, R.~Esmailzadeh,
S.~Dimopoulos:
 Phys.~Rev.~D {\bf 41}, 3594 (1990)

\bibitem{imp78} P.C.~McGuire, T.~Bowen,  R.~Mewaldt, P.~Steinhardt:
in preparation 
\bibitem{imax1} P.C.~McGuire and T.~Bowen: Proc.~5th October Astrophysics 
Conf.~in College Park, Maryland on Dark Matter, AIP, ed.~by S.S.~Holt and
C.L.~Bennett, Vol.~336, p.~53 (1994)
\bibitem{diss} P.C.~McGuire:  {\em Balloon-Borne Direct Search for Ionizing Massive Particles as a Component of the Dark Galactic
Halo Matter}, Ph.D.~dissertation, University of Arizona (1994)
\bibitem{imax2} P.C.~McGuire, T.~Bowen: in preparation 
\bibitem{xqc}  D.~McCammon {\em et al.}: NIM A {\bf 370 },
 266 (1996)
\bibitem{xqcdata}  D.~McCammon: University of Wisconsin,   private communication (2000)
\bibitem{forth} D.~McCammon {\it et al.}: in preparation
\bibitem{Mewaldt} R.~Mewaldt: California Institute of Technology, private communication (2000)
\end{thebibliography}
\end{document}